\documentclass[12pt,twoside]{amsart}

\usepackage{amssymb}

\newcommand{\ds}{\displaystyle}
\def\EXP{\textrm{{\large e}}}
\newcommand{\uop}{\mathbf{u}}
\newcommand{\vop}{\mathbf{v}}

\newcommand{\xop}{\mathbf{x}}
\newcommand{\pop}{\mathbf{p}}
\newcommand{\ii}{\mathsf{i}}
\begin{document}

\vspace{2cm}

\title[]{Quantization scheme for modular $q$-difference equations}%
\author{Sergey Sergeev}%
\address{Department of Theoretical Physics,
Research School of Physical Sciences nd Engineering, Australian
National University, Canberra ACT 0200, Australia}
\email{sergey.sergeev@anu.edu.au}

\thanks{This work was supported by the Australian Research Council and in part by the grant
CGP CRDF RM1-2334-MO-02}%

%\keywords{}%

%\date{September 15, 2003}

%-----------------------------------------------------
\begin{abstract}
Modular pairs of some second order $q$-difference equations are
considered. These equations may be interpreted as a quantum
mechanics of a sort of hyperelliptic pendulum. It is shown the
quantization of a spectrum may be provided by the condition of the
analyticity of the wave function. Baxter's $t-Q$ equations for the
quantum relativistic Toda chain in the ``strong coupling regime''
are related to the system considered, and the quantization
condition for $Q$-operator is also considered.
\end{abstract}

\maketitle

\section{Introduction}

The last few years in the theory of exactly integrable models
there appeared several important models with the local Weyl
algebra as the algebra of observables with the special values of
Weyl's parameter $q$ \cite{Faddeev,liouville,toda}
\begin{equation}\label{q-def}
q\;=\;\EXP^{2\pi\ii\,
b^2}\;,\;\;\;b\;=\;
\EXP^{\ii\theta}\;,\;\;\;0<\theta<\frac{\pi}{2}\;.
\end{equation}
Remarkable feature of this regime, called ``the strong coupling
regime'', is that the complex conjugation of $q$ is equivalent to
the Jacoby modular transform,
\begin{equation}\label{q-conj}
q^*\;=\;\EXP^{-2\pi\ii b^{-2}}\;.
\end{equation}
(In this paper we will use the star as the complex conjugation
sign.)

Let the coordinate $\xop$ and the momentum $\pop$, forming the
Heisenberg algebra
\begin{equation}\label{HA}
[\xop,\pop]\;=\;\frac{\ii}{2\pi}\;,
\end{equation}
are used for the definition of the Weyl pair $\uop,\vop$
\begin{equation}\label{weyl}
\uop\;=\;\EXP^{2\pi b \xop}\;,\;\;\;\vop\;=\;\EXP^{2\pi b
\pop}\;,\;\;\; \uop\;\vop\;=\;q\;\vop\;\uop\;,
\end{equation}
and the dual (i.e. conjugated) pair
\begin{equation}\label{weylc}
\uop^\dag\;=\;\EXP^{2\pi b^{-1}
\xop}\;,\;\;\;\vop^\dag\;=\;\EXP^{2\pi b^{-1} \pop}\;,\;\;\;
\vop^\dag\;\uop^\dag\;=\;q^*\;\uop^\dag\;\vop^\dag\;.
\end{equation}
Any elements of the first pair evidently commute with any element
of the second one. The physical meaning of the regime
(\ref{q-def}) is that any \emph{polynomial}
$\mathbf{J}(\uop,\vop)$ commutes with its conjugated,
\begin{equation}
\mathbf{J}(\uop,\vop)\;\mathbf{J}(\uop,\vop)^\dag\;=\;
\mathbf{J}(\uop,\vop)^\dag \mathbf{J}(\uop,\vop)\;,
\end{equation}
i.e. one may talk about commuting hermitian ``\emph{hamiltonians}''
$\mathbf{J}+\mathbf{J}^\dag$ and
$\ii(\mathbf{J}-\mathbf{J}^\dag)$.

Since $\mathbf{J}$ is a polynomial with initially arbitrary
coefficients, one may choose the spectral problem as follows,
\begin{equation}\label{spec-prob}
\langle\Psi|\,\cdot\,\mathbf{J}(\uop,\vop)\;=\;
\langle\Psi|\,\cdot\,\mathbf{J}(\uop,\vop)^\dag\;=\;0\;.
\end{equation}
These equations, being supplemented with an appropriate extra
condition for $\Psi$, should provide a quantization of
coefficients of polynomial $\mathbf{J}$. This quantization
condition and its realization as a set of well defined equations
for a particular $\mathbf{J}(\uop,\vop)$ is the subject of this
paper.

\section{Framework of quantum mechanics: $q$-hyperelliptic pendulum}

Here we will consider the following operator-valued polynomial
$\mathbf{J}(\uop,\vop)$:
\begin{equation}\label{Jqm}
\mathbf{J}(\uop,\vop)\;=\;\vop+\vop^{-1}+ T(\uop)\;,
\end{equation}
where the ``potential''
\begin{equation}\label{potential}
T(u)\;=\;\lambda u^{-L}\,\sum_{j=0}^{2L} t_{j} \cdot
(-u)^j\;,\;\;\;t_0=t_{2L}=1\;\;,
\end{equation}
and all $t_j$ and $\lambda$ are complex coefficients. This
$\mathbf{J}$ may be considered evidently as a $q$-deformation of
Schroedinger's kernel $\mathbf{j}=\pop^2+V(\xop)-E$, and $\Psi$ is
a wave function. Note, when $q\mapsto 1$, $J(u,v)=0$ defines the
genus $g=2L-1$ hyperelliptic curve with $g$ moduli $t_j$, and
mechanically the dynamic of $u$ and $v$ is a hyperlliptic
pendulum.

Let us choose the coordinate representation,
\begin{equation}\label{wf}
\langle\Psi| x\rangle\;=\;\Psi(x)\;,\;\;\;
\langle\Psi|\xop|x\rangle\;=\;x\Psi(x)\;,
\langle\Psi|\pop|x\rangle\;=\;\frac{\ii}{2\pi}\Psi'(x)\;.
\end{equation}
Therefore, due to (\ref{weyl}) and (\ref{weylc}),
\begin{equation}\label{repr}
\begin{array}{l} \ds
\langle\Psi|\uop|x\rangle\;=\;\EXP^{2\pi x b}\Psi(x)\;,\;\;\;
\langle\Psi|\uop^\dag|x\rangle\;=\;\EXP^{2\pi x b^{-1}}\Psi(x)\;,\\
\\
\langle\Psi|\vop|x\rangle\;=\;\Psi(x+\ii b)\;,\;\;\;
\langle\Psi|\vop^\dag|x\rangle\;=\;\Psi(x+\ii b^{-1})\;.
\end{array}
\end{equation}
For the shortness of all subsequent formulas, define the
correspondence
\begin{equation}\label{xbb}
\forall\;\;x\in\mathbb{C}\;\;\Leftrightarrow\;\; u\;\equiv\;
\EXP^{2\pi x
b}\;\;\;\textrm{and}\;\;\;\widetilde{u}\;\equiv\;\EXP^{2\pi x
b^{-1}}\;.
\end{equation}
Note, if $x$ is complex, $\widetilde{u}\neq u^*$.

Turn now to equations (\ref{spec-prob}) in this basis. For
(\ref{Jqm}) they are
\begin{equation}\label{diffeq}
\left\{
\begin{array}{l}
\Psi(x+\ii b)+\Psi(x-\ii b)\,+\,\Psi(x)\,T(u)\;=\;0\;,\\
\\
\Psi(x+\ii b^{-1})+\Psi(x-\ii b^{-1})
\,+\,\Psi(x)\,T^*(\widetilde{u})\;=\;0\;,
\end{array}\right.
\end{equation}
where $T^*$ means the complex conjugation of all coefficients of
the Laurent polynomial $T$. Function $\Psi(x)$ when
$x\in\mathbb{R}$ is the wave function, so that one condition for
$\Psi$ is evident:
\begin{equation}\label{L2}
\Psi(x)\;\in\;\mathbb{L}_2(\mathbb{R})\;.
\end{equation}
Clearly, the shifts of the argument of $\Psi$ to the complex
plane, corresponding to the second line of (\ref{repr}), are to be
understood as the complex continuation of the physical
$\Psi(x)_{x\in\mathbb{R}}$. In the framework of the quantum mechanics
the primitive way to generate the complex shifts is the series
expansion (recall, $\ii b=\ii \cos\theta -\sin\theta$),
\begin{equation}
\Psi(x+\ii b)\;=\;\sum_{n=0}^{\infty}\; \Psi^{(n)}(x-\sin\theta)\;
\frac{(\ii\cos\theta)^n}{n!}\;,
\end{equation}
so that (\ref{diffeq}) with $x\in\mathbb{R}$ is the coordinate
representation of (\ref{spec-prob}) only if $\Psi(x)$ is
analytical in the strip
\begin{equation}\label{strip}
-\cos\theta\;<\;\Im(x)\;<\;\cos\theta\;.
\end{equation}
Further we will see, this condition will provide the analyticity
of $\Psi(x)$ in the whole complex plane.

Note, system (\ref{diffeq}) is not related to the well known
theory of Harper's equations because of our $|q|\,\neq\,1$.

\section{Holomorphic counterparts}

The first key observation is that each equation of (\ref{diffeq})
may be modified to special forms, having the holomorphic on
$u^{\pm 1}$ or on $\widetilde{u}^{\pm 1}$ solutions. Let
\begin{equation}
t_+(u)=\frac{u^L}{\lambda}T(u)=\sum_{j=0}^{2L}
t_j\,\cdot\,(-u)^j\;,\;\;\;
t_-(u)=\frac{1}{u^L\lambda}T(u)=\sum_{j=0}^{2L}
t_{2L-j}\,\cdot\,(-u)^{-j}\;,
\end{equation}
and consider two equations
\begin{equation}\label{+}
\chi_+(q^{-1}u)= t_+(u)\chi_+(u) - G u^{2L}\chi_+(qu)
\end{equation}
and
\begin{equation}\label{-}
\chi_-(qu)= t_-(u)\chi_-(u) - G u^{-2L}\chi_-(q^{-1}u)\;,
\end{equation}
where
\begin{equation}\label{G}
G\;\equiv\;q^{L}\lambda^{-2}\;.
\end{equation}
Let $\chi_+(u)$ be the holomorphic on $u$ solution of (\ref{+}),
normalized as $\chi_+(0)=1$. Let $\chi_-(u)$ be the holomorphic on
$u^{-1}$ solution of (\ref{-}), normalized as $\chi_-(\infty)=1$.
Both these functions are uniquely defined. Expansion of e.g.
$\chi_+$ may be estimated
\begin{equation}
\chi_+(u)=\sum_{n=0}^\infty \chi_n\,\cdot\,(-u)^n\;,\;\;\;
|\chi_n|\;<\;q^{n^2/4L} X^n\;,
\end{equation}
where $X$ is defined by the set of $t_j$ and $G$. Besides, there
exists a way to represent $\chi_+$ as the semi-infinite
\emph{matrix} product. Let
\begin{equation}
L(u)\;=\;\left(\begin{array}{cc}
\ds t_+(u) & \ds -\, G\, u^{2L}\\
&\\
\ds 1 & \ds 0
\end{array}\right)
\end{equation}
and
\begin{equation}
L_n(u)\;=\; L(u)\cdot L(qu) \cdot L(q^2u) \cdots L(q^{n-1}u)\;.
\end{equation}
For $|q|<1$ the matrix function $L_n(u)$ is absolutely convergent
product, and
\begin{equation}
L_\infty(u)\;=\;\left(\begin{array}{cc}
\ds \chi_+(q^{-1}u) & \ds 0\\
&\\
\ds \chi_+(u) & \ds 0
\end{array}\right)\;.
\end{equation}
This form of $L_\infty$ follows from the relation
$L_\infty(u)=L(u)\,\cdot\,L_\infty(qu)$.

Analogous expressions may be written and for $\chi_-$. For
example, if $t_-(u)=t_+(u^{-1})$ (i.e. $T(u)=T(u^{-1})$), then
$\chi_-(u)\equiv\chi_+(u^{-1})$.

Function $\ds\psi(u)=\prod_{n=0}^\infty (1-q^nu)$ is called
sometimes as ``quantum dilogarithm'' since in the limit $\ds
q=\EXP^{-\epsilon}\mapsto 1$ the leading asymptotic is $\ds
\psi(u)=\exp\{\epsilon^{-1}\int_0^u \log(1-x) d\log x +O(1)\}$. In
the same limit holomorphic solution of
\begin{equation}
\chi(q^{-1}u)=t(u)\chi(u)-p(u)\chi(qu)\;,
\end{equation}
where polynomials $t(u)$ and $p(u)$ are normalized as $t(0)=1$ and
$p(0)=0$ (cf. (\ref{+})), has the asymptotic
\begin{equation}\label{asmpt}
\chi(u)\;=\;\exp\{\epsilon^{-1}\int_0^u\log y d\log x +O(1)\}
\end{equation}
where $x,y$ belongs to the hyperelliptic curve
\begin{equation}
y+p(x)y^{-1}\;=\;t(x)\;.
\end{equation}
Initial point in the integral (\ref{asmpt}) is the point $x=0,y=1$
of the curve, integration is performed in the neighborhood of this
point.

\section{Ansatz for the wave function}

With the definition (\ref{xbb}) taken into account, both
\begin{equation}\label{psipm}
\psi_+(x)=\frac{\chi_+(u)}{n_+(u)}\;\;\;\textrm{and}\;\;\;
\psi_-(x)=\frac{\chi_-(u)}{n_-(u)}
\end{equation}
solve the first equation of (\ref{diffeq}) if
\begin{equation}
\frac{n_+(u)}{n_+(q^{-1}u)}\;=\;-\lambda u^{-L}\;,\;\;\;
\frac{n_-(u)}{n_-(qu)}\;=\;-\lambda u^L\;.
\end{equation}
Solutions of the second equation of (\ref{diffeq}) are simply
\begin{equation}
\psi_+^*(x)=\frac{\chi_+^*(\widetilde{u})}{n_+^*(\widetilde{u})}\;\;\;\textrm{and}\;\;\;
\psi_-^*(x)=\frac{\chi_-^*(\widetilde{u})}{n_-^*(\widetilde{u})}\;.
\end{equation}
Then a general solution of (\ref{diffeq}) is a linear combination
of four terms,
\begin{equation}
\Psi(x)\;\sim\;\sum_{\varepsilon,\varepsilon'=\pm}
\psi_\varepsilon^{}(u)\psi_{\varepsilon'}^*(\widetilde{u})
\end{equation}
One may estimate the average asymptotic of any
$\psi_\varepsilon^{}(u)\psi_{\varepsilon'}^*(\widetilde{u})$
\footnote{Corollary: Let $f(x)$ has the asymptotic difference
properties
\begin{equation}
\frac{f(x)}{f(x+\ii b)}\;\sim\;a\EXP^{2\pi x b
\varepsilon}\;,\;\;\; \frac{f(x)}{f(x -\ii
b^{-1})}\;\sim\;a'\EXP^{2\pi x b^{-1} \varepsilon'}
\end{equation}
when $x$ is big, $\varepsilon$ and $\varepsilon'$ are integers.
Then the average asymptotic of $|f(x)|$ with real
$x\mapsto\pm\infty$ is
\begin{equation}\label{ass}
|f(x)|\;\sim\;\exp\left(\frac{\pi}{2}(\varepsilon+\varepsilon')\cot\theta
x^2 + \pi(\varepsilon+\varepsilon')\cos\theta x +
\frac{\log|aa'|}{2\sin\theta} x\right)\;.
\end{equation}
} and conclude that $\psi_\pm^{}(u)\psi_{\pm}^*(\widetilde{u})$
grow at $x\mapsto\pm\infty$, while
\begin{equation}
|\psi_\pm^{}(u)\psi_{\mp}^*(\widetilde{u})|\;\sim\;\EXP^{-2\pi L
\cos\theta |x|}\;,\;\;\;x\mapsto\pm\infty
\end{equation}
Thus we come to the following ansatz for
$\Psi(x)\in\mathbb{L}_2(\mathbb{R})$:
\begin{equation}\label{a2}
\Psi(x)\;=\;\frac{\chi_+^{}(u)\chi_-^*(\widetilde{u})}{D_1(x)}\;-\;
\frac{\chi_-^{}(u)\chi_+^*(\widetilde{u})}{D_2(x)}
\end{equation}
where $D_1\sim n_+^{}n_-^*$ and $D_2\sim n_-^{}n_+^*$ obey
\begin{equation}\label{D12}
\begin{array}{ll}
\ds\frac{D_1(x)}{D_1(x-\ii b)}\;=\;-\lambda u^{-L}\;,&
\ds\frac{D_1(x)}{D_1(x-\ii b^{-1})}\;=\;-\lambda^* \widetilde{u}^L\\
&\\
\ds\frac{D_2(x)}{D_2(x+\ii b)}\;=\;-\lambda u^L\;,&
\ds\frac{D_2(x)}{D_2(x+\ii b^{-1})}=-\lambda^*
\widetilde{u}^{-L}\;.
\end{array}
\end{equation}
Suppose further
\begin{equation}\label{G-gamma}
-\lambda\;=\;\EXP^{\ii\pi L b^2 -\pi\gamma b}\;,\;\;\;
G\;=\;\EXP^{2\pi\gamma b}
\end{equation}
with $\gamma\in\mathbb{R}$. Then $D_1$ and $D_2$ may be specified
as follows:
\begin{equation}
\begin{array}{l}
\ds D_1(x)\;=\;\EXP^{-\ii\pi L x^2 + \ii\pi\gamma} u^{-L}
W_1(u)\;,\\
\\
\ds D_2(x)\;=\;\EXP^{-\ii\pi L x^2 - \ii\pi\gamma} u^{-L}
W_2(u)\;,
\end{array}
\end{equation}
where
\begin{equation}\label{W12}
W_{1,2}(u)\;=\; u^{2L} W_{1,2}(qu)\;.
\end{equation}
Functions $D_1$ and $D_2$, and correspondingly $W_1$ and $W_2$ may
be defined via (\ref{W12}) up to two arbitrary double periodic
functions. Equation (\ref{W12}) implies that $W_{1,2}(u)$ as
functions of $x$ has at least $2L$ zeros in the parallelogram of
periods. In general, the zeros may be chosen to be complex, so
that such $\Psi(x)\in\mathbb{L}_2(\mathbb{R})$ and nothing has
been quantized. Therefore one should turn to the second condition
for $\Psi(x)$, the analyticity in the strip (\ref{strip}).

The strip (\ref{strip}) contains the whole parallelogram of
periods, so that there exists only one way to provide the
analyticity of $\Psi(x)$ in the strip. This way is the following:
\begin{itemize}
\item at the first, $W_2$ should be chosen proportional to $W_1$ so
that they give the common denominator in (\ref{a2}),
\begin{equation}\label{a3}
\Psi(x)\;=\;\EXP^{\ii\pi L x^2}\frac{\EXP^{-\ii\pi\gamma
x}\chi_+^{}(u)\chi_-^*(\widetilde{u})-\xi\EXP^{\ii\pi\gamma
x}\chi_-^{}(u)\chi_+^*(\widetilde{u})}{u^{-L}W(u)}\;,
\end{equation}
where $\xi$ is a complex number.
\item secondly $W(u)$ should be chosen with the minimal set of zeros,
\begin{equation}
W(u)\;=\;\prod_{j=1}^{2L} H(u/s_j)\;,
\end{equation}
where the theta-function $H(u)=-uH(qu)$ is given by
\begin{equation}
H(u)\;=\;(u;q)_\infty
(qu^{-1};q)_\infty(q;q)_\infty=\sum_{n\in\mathbb{Z}} q^{n(n-1)/2}
(-u)^n\;,
\end{equation}
so that (\ref{W12}) provides
\begin{equation}
\prod_{j=1}^{2L}\;s_j\;=\;1\;,
\end{equation}
\item
and thirdly, one should provide the Gutzwiller principle
\cite{Gutzwiller}: the cancellation of zeros of $W(u)$ and zeros
of the numerator of (\ref{a3}) in the whole strip.
\end{itemize}

\section{Quantization equations}

Let us in addition to the conventions (\ref{xbb}) and (\ref{G-gamma}) use
\begin{equation}\label{s-sigma}
s_j=\EXP^{2\pi \sigma_j
b}\;,\;\;\;\sum_{j=1}^{2L}\sigma_j\;=\;0\;.
\end{equation}
Zeros of denominator of (\ref{a3}) in the strip (\ref{strip}) are
$x=\sigma_j+\ii n b - \ii n b^{-1}$, $j=1,...,2L$,
$n\in\mathbb{Z}$. This corresponds to $u=q^n s_j$ and
$\widetilde{u}=q^{*n}\widetilde{s_j}$. Conditions of zero value of
the numerator of (\ref{a3}) in this points are the infinite set of
equations
\begin{equation}\label{infset}
\xi\;=\;\EXP^{-2\pi\ii\gamma\sigma_j}\; \frac{\ds
G^n\frac{\chi_+^{}(q^n s_j)}{\chi_-(q^n s_j)}}{\ds G^{*n}
\frac{\chi_+^*(q^{*n}\widetilde{s_j})}{\chi_-^*(q^{*n}\widetilde{s_j})}}
\end{equation}
There exists the unique way to satisfy all this infinite set. At
the first, one has to demand that both numerator and denominator
of (\ref{infset}) do not depend on $n$, i.e.
\begin{equation}\label{M}
G^n\frac{\chi_+^{}(q^n s_j)}{\chi_-(q^n s_j)}\;=\;
\frac{\chi_+^{}(s_j)}{\chi_-^{}(s_j)}\;
\stackrel{\textrm{def}}{=}\;M(s_j)\;,
\end{equation}
The special set of $s_j$, obeying this relations, corresponds to
the zeros of the $q$-Wronskian of $\psi_+$ and $\psi_-$,
(\ref{psipm}). Excluding $n_\pm$, we define $q$-Wronskian as
\begin{equation}\label{W}
W(u)\;\stackrel{\textrm{def}}{=}\;
\chi_+(q^{-1}u)\,\chi_-(u)\,-\,G\, \chi_+(u)\,\chi_-(q^{-1}u)\;.
\end{equation}
Due to the equations for $\chi_\pm$, (\ref{+},\ref{-}),
\begin{equation}
W(u)\;=\;u^{2L}\,W(qu)\;,
\end{equation}
and since (\ref{W}) by the definition is a convergent series in
$u,u^{-1}$, there exists the unique  set of $s_j$, $j=1,...,2L$, such
that
\begin{equation}\label{wro}
W(u)\;=\;C\,\prod_{j=1}^{2L}\;
H(u/s_j)\;,\;\;\;\prod_{j=1}^{2L}\;s_j\;=\;1\;,
\end{equation}
A constant $C$, depending on $G$ and the set of $t_j$, is
unessential. The periodicity of the Wronskian (\ref{W}) provides
(\ref{M}).

Therefore, if $s_j$ are the zeros of the Wronskian (\ref{W}), all
the infinite set (\ref{infset}) reduces to $2N-1$ equations (since
$\xi$ itself is to be defined by these equations)
\begin{equation}\label{qeqs}
\xi\;=\;\EXP^{-2\pi\ii\gamma\sigma_j}\;
\frac{M(s_j)}{M^*(\widetilde{s_j})}\;,\;\;\;j=1,...,2L\;.
\end{equation}
It is the set of equations for the coefficients $t_j$ of the
``potential'' (\ref{potential}), the number of equations
corresponds exactly to the number of unknown $t_j$ and to the
number of independent $s_j$. Due to the definition
(\ref{W},\ref{wro}), the set $s_j$ is the unique function of
$t_j$. From the other side, one may show that $t_j$ are
multivalued functions of $s_j$ with the infinite number of leafs,
this provides a discrete infinite set of solutions of
(\ref{qeqs}).

A numerical investigation of (\ref{qeqs}) allows us to assume that
all solutions of (\ref{qeqs}) correspond to real $\sigma_j$, so
that $\xi$ is a pure phase. Besides, it provides that $\Psi(x)$
(\ref{a3}) has a constant phase when $x\in\mathbb{R}$ (since
complex conjugation is equivalent to the modular transform). Also,
if the ``potential'' is symmetrical,  $T(u)=T(u^{-1})$, then
$\xi=\pm 1$ is the parity.

Note finally, due to the definition of $M$, (\ref{M}), the
numerator of (\ref{a3}) is zero in the points $x=\sigma_j+\ii n b
- \ii m b^{-1}$, $n,m\in\mathbb{Z}$, therefore $\Psi(x)$ is
analytical in the whole complex plane.

\section{Quantum relativistic Toda chain}

The quantum curve for the length $N$ quantum relativistic Toda
chain in one of possible normalizations \cite{qrtc} is
\begin{equation}
\mathbf{J}(\uop,\vop)\;=\; \vop\;+\;(-)^N G \uop^N \vop^{-1} \;-\;
t(\uop)\;,
\end{equation}
where
\begin{equation}
t(u)=\sum_{j=0}^{N} t_j\,\cdot\,(-u)^j\;,\;\;\;t_0=t_N=1\;.
\end{equation}
Baxter's equations for $Q(x)=\langle Q|x\rangle$ are $\langle
Q|\,\mathbf{J}=\langle Q|\,\mathbf{J}^\dag=0$.

The conditions for $Q(x)$ are the following \cite{toda,Smirnov}:
\begin{itemize}
\item $|Q(x)|\;\sim\;1$ when $x\mapsto -\infty$ and
$|Q(x)|\;\sim\;\EXP^{-2\pi N \cos\theta x}$ when $x\mapsto
+\infty$, and
\item $Q(x)$ is analytical in the whole complex plane.
\end{itemize}

Then, just modifying slightly all the previous considerations
(they correspond actually to even $N=2L$), at the first define
$\chi_\pm$ as holomorphic on $u^{\pm 1}$ solutions of
\begin{equation}
\begin{array}{l}
\ds \chi_+^{}(q^{-1}u)\;=\;t(u)\,\chi_+^{}(u)\,-\,G\,(-u)^N
\chi_+^{}(qu)\;,\\
\\
\ds
\chi_-^{}(qu)\;=\;\frac{t(u)}{(-u)^N}\,\chi_-^{}(u)\,-\,\frac{G}{(-u)^{N}}\,
\chi_-^{}(q^{-1}u)\;.
\end{array}
\end{equation}
Formula (\ref{W}) for their $q$-Wronskian is the same, but
$W(u)=(-u)^N W(qu)$ and therefore the decomposition is
\begin{equation}
W(u)\;=\;C\;\prod_{j=1}^{N}\;H(u/s_j)\;,\;\;\;\prod_{j=1}^N\;s_j=1\;.
\end{equation}
As previously, let
\begin{equation}
M(s_j)\;=\;\frac{\chi_+(s_j)}{\chi_-(s_j)}\;,
\end{equation}
and the quantization condition for real $\sigma_j$,
$s_j\equiv\EXP^{2\pi\sigma_j b}$, is
\begin{equation}
\EXP^{-2\pi\ii\gamma\sigma_j}\;\frac{M(s_j)}{M(s_j)^*}\;=\;
\xi\;,\;\;\;j=1,...,N\;.
\end{equation}
All notations here are the same as in
(\ref{G-gamma},\ref{s-sigma}, etc.)

Now $Q(x)$, obeying all the conditions, is given by
\begin{equation}
Q(x)\;=\;\frac{\EXP^{-2\pi\ii\gamma
x}\chi_+^{}(u)\chi_-^*(\widetilde{u})-\xi\chi_-^{}(u)\chi_+^*(\widetilde{u})}{W(u)}\;.
\end{equation}

\section{Conclusion}

In this paper we have proposed the quantization scenario for some
Baxter-type equations  in the strong coupling regime. The
important observation has been made that certain second-order
$q$-difference equations may have holomorphic on $u$ solutions
(functions $\chi$) in the same way as the first order difference
equations have the holomorphic solution -- the compact quantum
dilogarithm \cite{fk-qd}. These $\chi$ -- functions (they are
related to the Hill determinants in the usual quantum Toda chain)
are suitable for numerical analysis, contrary to the case $|q|=1$
\cite{Smirnov}. Another important observation has been made that
the zeros of $q$-Wronskian of two $\chi$-functions reduces the
infinite set of residue relations to the finite one.

Quantization condition, the analyticity of the wave function or of
the Baxter function $Q$ in the strip (\ref{strip}), is the set of
transcendental equations (\ref{qeqs}) in the terms of the
functions $\chi$. The number of equations (\ref{qeqs}) grows when
the size of the system grows. We still did not find a good method
of investigation of these equations in the thermodynamical limit.

In this paper we have dealt with rather specific quantum curves
$\mathbf{J}(\uop,\vop)$. In the both examples $\Psi(x)$ and $Q(x)$
are analytical in the whole complex plane, this is the specific
feature of Toda-type $\mathbf{J}(\uop,\vop)$. For more complicated
examples (e.g. an appropriately reformulated sine-Gordon model) it
will not be so. We believe, the scheme proposed here, may be
generalized to the class of all quantum algebraic curves (not only
hyperelliptic), corresponding to the class of integrable models
with the local Weyl algebra as the algebra of observables
\cite{Sergeev}. This will be the subject of forthcoming papers.

\noindent\textbf{Acknowledgements} I would like to thank V.
Bazhanov, R. Kashaev and S. Pakuliak for many useful discussions.

\end{document}